\documentclass[preprint,11pt]{aastex}
\usepackage{epsfig}
\usepackage{rotating}      
\tightenlines

\def	\Angstrom	{\,{\rm \AA}}
\def	\beq	{\begin{equation}}
\def	\eeq	{\end{equation}}
\def	\gpar	{{g}}
\def	\gtsim	{\gtrsim}		 
\def	\ltsim	{\lesssim}		 

\newlength{\figwidth}
\newlength{\figwidthland}
\newlength{\figwidthport}
\newlength{\figwidthportb}
\countdef\decade=200
\advance\decade by \year
\countdef\hours=201
\advance\hours by \time
\divide\hours by 60
\countdef\mins=202
\advance\mins by \hours
\multiply\mins by 60
\multiply\hours by 100
\countdef\miltime=203
\advance\miltime by \hours
\advance\miltime by \time
\advance\miltime by -\mins

\begin{document}

\title{
        \vspace*{-3.0em}
        {\normalsize\rm To appear in {\it The Astrophysical Journal}}\\ 
        \vspace*{1.0em}
Scattering by Interstellar Dust Grains.
I. Optical and Ultraviolet
	}

\author{B.T. Draine			
	}
\affil{Princeton University Observatory, Peyton Hall, Princeton,
NJ 08544; \\
{\tt draine@astro.princeton.edu}}

\begin{abstract}
Scattering and absorption properties at optical and ultraviolet
wavelengths are calculated
for an interstellar dust model consisting of carbonaceous grains
and amorphous silicate grains.
Polarization as a function of scattering angle is calculated
for selected wavelengths from the infrared to the
vacuum ultraviolet.

The widely-used Henyey-Greenstein phase function provides
a good approximation for the scattering phase function at
wavelengths between $\sim$0.4 and 1$\micron$,
but fails to fit the calculated phase functions at shorter and longer
wavelengths.
A new analytic phase function is presented.
It is exact at long wavelengths,
and provides a good
fit to the numerically-calculated phase function for
$\lambda > 0.27\micron$.

Observational determinations of the scattering albedo and
$\langle\cos\theta\rangle$ show considerable disagreement, especially
in the ultraviolet.  Possible reasons for this are discussed.
\end{abstract}


\section{Introduction
	\label{sec:intro}}

Interstellar grains scatter electromagnetic radiation.
Reflection nebulosities are conspicuous at optical and UV 
wavelengths when dust is brightly illuminated by a nearby star.
Dust clouds which are not unusually close to a star are illuminated by
the general interstellar radiation field.
Finally, the starlight scattered by dust in the diffuse interstellar medium
constitutes the so-called ``diffuse galactic light''.
Observations of reflection nebulae, dust clouds, 
and the diffuse galactic light,
provide a means of determining 
the scattering properties of interstellar grains, thereby testing models
for interstellar dust.

Photoionization and photodissociation of molecules play a major
role in interstellar chemistry, and the
chemical structure of molecular clouds is therefore directly linked to the
ability of ultraviolet starlight to penetrate into these dusty regions.
Knowledge of dust scattering properties in the ultraviolet is therefore
required for realistic modeling of interstellar clouds.

The nature of interstellar grains remains uncertain
(see Draine 2003a, and references therein).  
This paper will examine
the scattering properties for a grain model consisting of two
separate grain populations -- carbonaceous grains and silicate grains.
With the grains approximated by homogeneous spheres with the
size distributions found by Weingartner \& Draine (2001;
hereafter WD01), this grain model is consistent with the observed
interstellar extinction, 
the observed infrared emission from interstellar dust 
(Li \& Draine 2001, 2002),
and the
X-ray scattering halo observed around Nova Cygni 1992
(Draine \& Tan 2003).
The carbonaceous grains are assumed to be primarily carbon when the grains
are large, but to extend down to very small sizes with the smallest
grains being individual polycyclic aromatic hydrocarbon molecules.
The scattering is dominated by grains with radii $a\gtsim 100\Angstrom$,
containing $\gtsim 10^6$ atoms;
carbonaceous grains in this size range are modeled using the optical
properties of graphite.

The primary
objective of this paper is to calculate the scattering and extinction
properties of this dust model at infrared, optical, and ultraviolet
wavelengths, and to make these results available for
use in radiative transfer calculations and for comparison with observations.
The X-ray scattering and absorption properties of this grain model
are the subject of Paper II (Draine 2003b).

The adopted dielectric functions are presented in \S\ref{sec:dielec}.
The scattering properties of
interstellar dust at optical and ultraviolet energies, as calculated for
the carbonaceous-silicate grain model,
are presented in \S\ref{sec:optuv_scat}.
We show the scattering phase
function at selected wavelengths from the SDSS z band ($\lambda=8930\Angstrom$)
to the vacuum ultraviolet ($\lambda=1820\Angstrom$).
Scattering properties are calculated for Milky Way dust with
$R_V=3.1$, and also for models for dust in the LMC and SMC.
In \S\ref{sec:pol} we show the degree of polarization as a function of
scattering angle for selected wavelengths.

The Henyey-Greenstein phase function has often been used to approximate the
anisotropic scattering properties of interstellar dust.  
In \S\ref{sec:HG} we show that the Henyey-Greenstein function
has r.m.s. error $<10\%$
for $0.47\micron < \lambda < 0.94\micron$, but
has larger errors outside this range.
We present a new analytic phase function (equation \ref{eq:newPhi})
with a wider range of applicability,
with r.m.s. error $<10\%$ for 
$\lambda > 0.27\micron$.

In \S\ref{sec:discuss} we collect observational determinations of the
scattering albedo and $\langle\cos\theta\rangle$ for dust in
reflection nebulae, in dense clouds, and in the diffuse interstellar medium.
Discrepancies among these determinations are noted, and possible reasons
for this are discussed.

The principal results are summarized in \S\ref{sec:summary}

\section{Dielectric Function\label{sec:dielec}}

As discussed by WD01, the grain population is assumed to
include very small grains with the optical properties
of polycyclic aromatic hydrocarbon molecules (PAHs), 
plus larger grains which are approximated as carbonaceous or
silicate spheres.
The PAHs produce negligible scattering.

From the observed 3.4$\micron$ C-H stretch feature,
Pendleton \& Allamandola (2002) estimate that $\sim$85\% of the C is 
aromatic, and $\sim$15\% is aliphatic (chainlike).
The graphite dielectric function will be used to approximate the optical
and ultraviolet response of interstellar carbonaceous grain material.
Scattering and absorption by the carbonaceous spheres is calculated using
the dielectric tensor of graphite, using the usual 
``1/3-2/3 approximation'' (Draine \& Malhotra 1993).

\begin{figure}[h]
\centerline{\epsfig{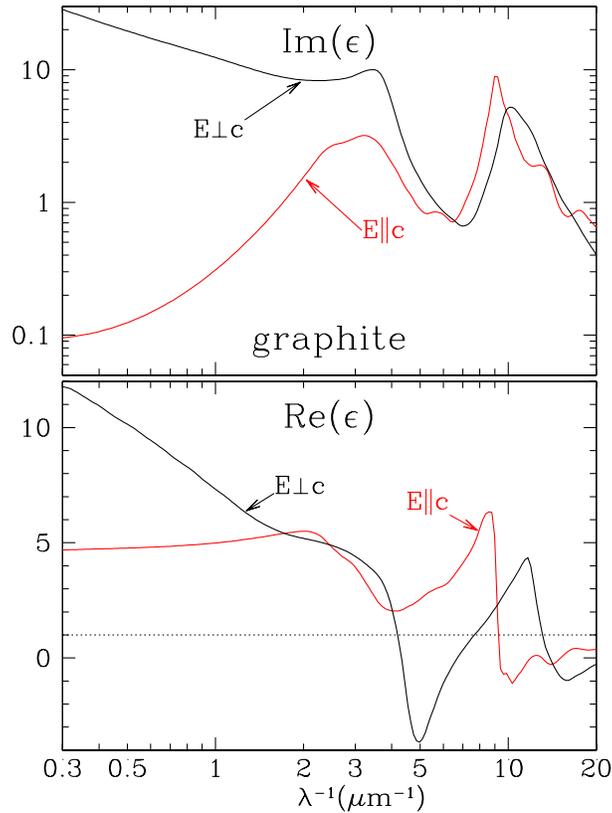}}
\caption{\label{fig:eps_C}\footnotesize
	Dielectric function for graphite 
	(Draine 2003b).
	}
\end{figure}

\begin{figure}[h]
\centerline{\epsfig{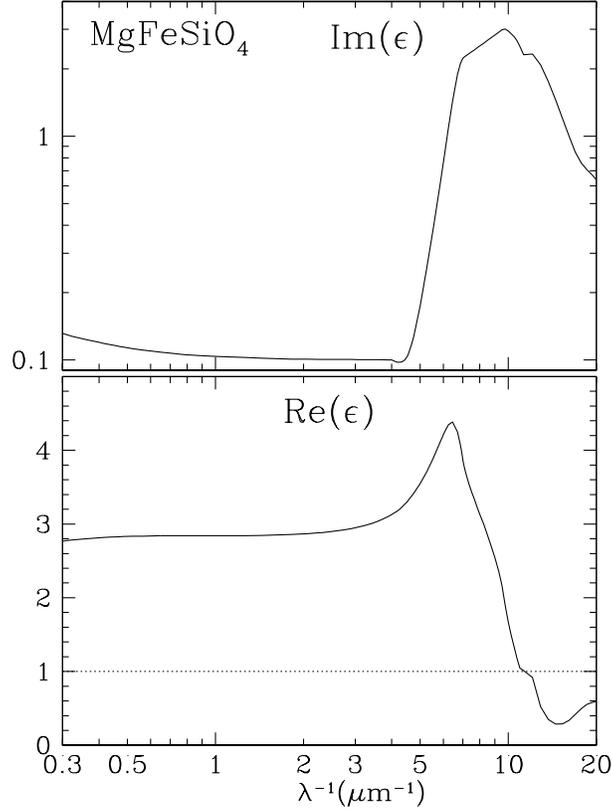}}
\caption{\label{fig:eps_sil}\footnotesize
	Dielectric function adopted for amorphous MgFeSiO$_4$
	(Draine 2003b).}
\end{figure}

The dielectric functions used here are taken from Paper II, which
constructs self-consistent dielectric functions extending from microwave
to X-ray energies, including realistic structure near X-ray absorption
edges.
The adopted dielectric functions for graphite and ``astronomical
silicate'' are shown in Figures
\ref{fig:eps_C} 
and \ref{fig:eps_sil}.
These dielectric functions are close to the dielectric functions obtained
previously by Draine \& Lee (1984), although differing in detail.

\section{IR-Optical-UV Scattering by Interstellar Dust\label{sec:optuv_scat}}

\subsection{Angular Distribution of Scattered Light}

Weingartner \& Draine (2001; hereafter WD01) obtained size distributions
of spherical
carbonaceous and silicate grains which reproduce
the observed extinction curve both in the local Milky Way and in the
Large and Small Magellanic Clouds.
Here we calculate the scattering properties of these dust mixtures.

The scattering properties of a particular dust mixture 
$X$ are characterized by the differential scattering cross section
per H nucleon
\beq
\left(\frac{d\sigma_{\rm sca}(\lambda,\theta)}{d\Omega}\right)_X \equiv 
\sum_j \int da \left(\frac{1}{n_{\rm H}} \frac{dn_j}{da}\right)_X
\left(\frac{dC_{\rm sca}}{d\Omega}\right)_{j,a,\lambda}
~~~,
\eeq
where $n_{\rm H}^{-1} (dn_j/da)da$ is the number of grains of type $j$ per
H nucleon with radii in $(a,a+da)$, and $(dC_{\rm sca}/d\Omega)_{j,a,\lambda}$
is the differential scattering cross section for grain type $j$,
radius $a$,
at wavelength $\lambda$, for scattering angle $\theta$, for a grain
illuminated by unpolarized light.
The grains are assumed to be spherical, and the differential scattering
cross sections are calculated using Mie theory (Bohren \& Huffman 1984),
using the code developed by Wiscombe (1980, 1996).

\begin{figure}[h]
\centerline{\epsfig{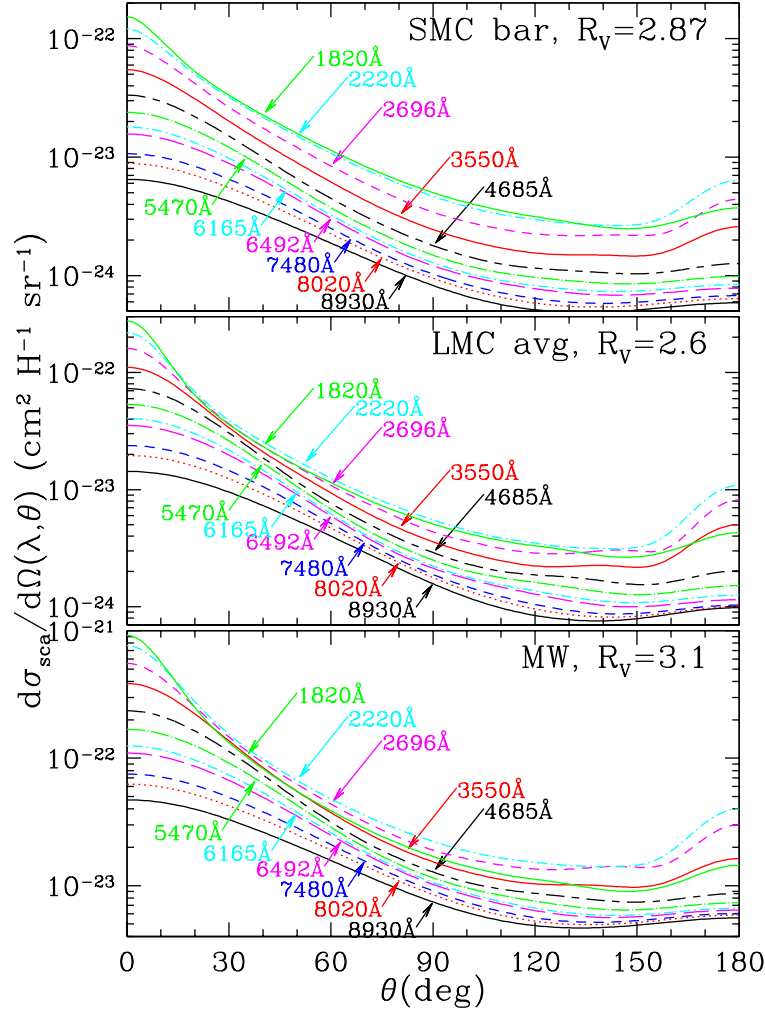}}
\caption{\footnotesize
	\label{fig:F_X}
	$d\sigma_{\rm sca}/d\Omega$ at
	selected wavelengths $\lambda$, as a function of the
	scattering angle $\theta$,
	for WD01 models for Milky Way dust with $R_V=3.1$, LMC average dust,
	and SMC bar dust.
	Curves are labelled by wavelength $\lambda$.
	}
\end{figure}

In Figure \ref{fig:F_X} we show the differential scattering cross section
per H nucleon for $R_V=3.1$ Milky Way dust, at 
the central wavelengths of SDSS 
z~(8930\AA), i~(7480\AA),
r~(6165\AA), g~(4685\AA), and u~(3550\AA),
Cousins I~(8020\AA) and R~(6492\AA), V~(5470\AA),
and the F250W~(2696\AA), F220W~(2220\AA), and F25CN182~(1820\AA)  
filters for the Space Telescope Imaging Spectrograph (STIS).
The scattering becomes stronger and more forward-throwing at shorter 
wavelengths.

\begin{figure}[h]
\centerline{\epsfig{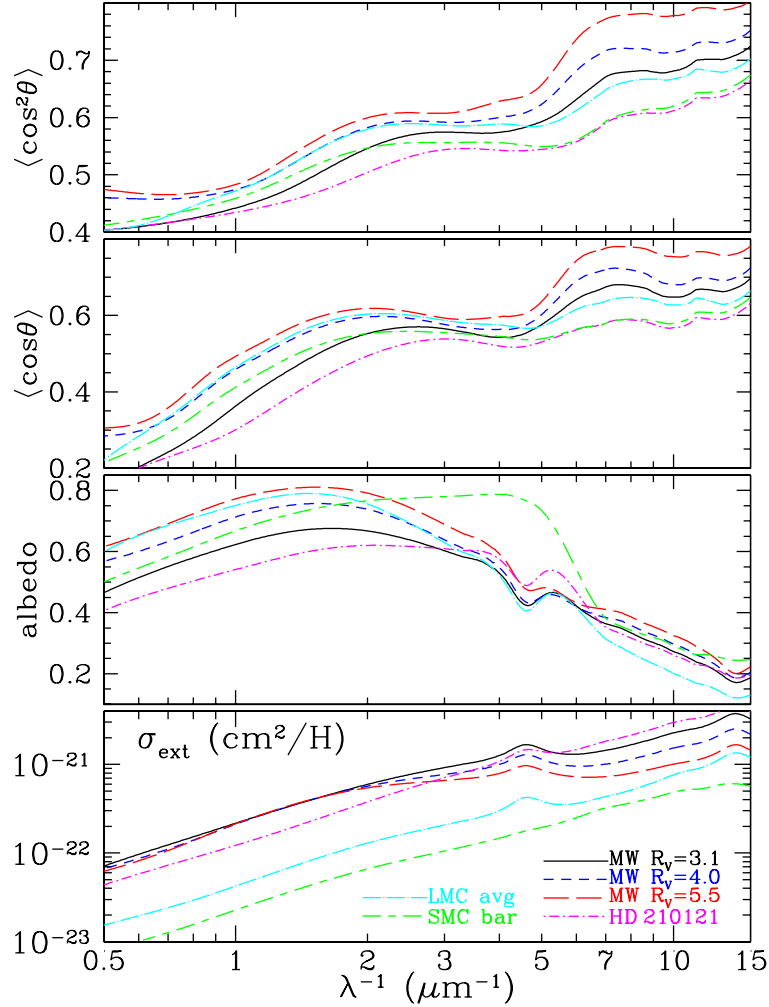}}
\caption{\footnotesize
	\label{fig:sigext,albedo,cos,cos2}
	Extinction cross section per H, albedo, $\langle\cos\rangle$,
	and $\langle\cos^2\rangle$ calculated for WD01 models
	for Milky Way dust with $R_V=3.1$, 4.0, and 5.5,
	average LMC dust, and dust in the SMC bar.
	}
\end{figure}
To see the sensitivity to variations in the dust mixture, Fig.\ \ref{fig:F_X}
also
shows $d\sigma_{\rm sca}/d\Omega$ 
calculated for the WD01 dust mixtures for
the ``average LMC'' and the SMC bar.
The calculated scattering
closely resembles the Milky Way scattering, but with an overall reduction of
about a factor of 4 for the LMC, and a factor of 8 for the SMC, 
in line with the overall dust and metal abundance
relative to the Milky Way.\footnote{
	Ne/H is $\sim30\%$ of solar in the LMC and $\sim14\%$
	of solar in the SMC
	(Dufour 1984; Kurt \& Dufour 1998).
	In the LMC, $E(B-V)/N_{\rm H}$ is $\sim24\%$
	of the local Milky Way value (Koorneef 1982; Fitzpatrick 1986)
	while in the SMC it is only $\sim13\%$ of the local value
	(Martin et al. 1989).%
	}

Figure \ref{fig:sigext,albedo,cos,cos2} shows the wavelength dependence of 
$\sigma_{\rm ext}$ (the total extinction cross section per H), the scattering
albedo $\sigma_{\rm sca}/\sigma_{\rm ext}$, and the 
first and second moments\footnote{%
	The second moment $\langle\cos^2\rangle$ will be used in
	the new analytic phase function proposed in \S\ref{sec:new_phase_func}.
	}
$\langle\cos\theta\rangle$ and $\langle\cos^2\theta\rangle$ for the scattered
radiation.
Six different grain models are shown,
fitted to different observed extinction curves (see WD01 for details).
There are considerable differences in albedo, $\langle\cos\rangle$,
and $\langle\cos^2\rangle$ among the models.
In particular, the SMC bar model shows a high albedo near $4.6\micron^{-1}$,
in contrast to the other five models which have a local minimum in the
albedo at this frequency.
This is because the SMC bar model differs from the other 5 models in
lacking PAHs and $a\ltsim0.02\micron$ graphite grains, as these are
not allowed by the observed absence of a $4.6\micron^{-1}$ extinction
``bump''.

\subsection{Polarization of Scattered Light\label{sec:pol}}

Even when a grain is illuminated by unpolarized light, the scattered
radiation is generally polarized.
The degree of polarization depends upon both the scattering angle
and the wavelength of the radiation.
The fractional polarization 
$P\equiv (I_\perp-I_\parallel)/(I_\perp+I_\parallel)$, where
$I_\perp, I_\parallel$ are the intensities of scattered light in polarization
modes perpendicular or parallel to the scattering plane.

\begin{figure}[h]
\centerline{\epsfig{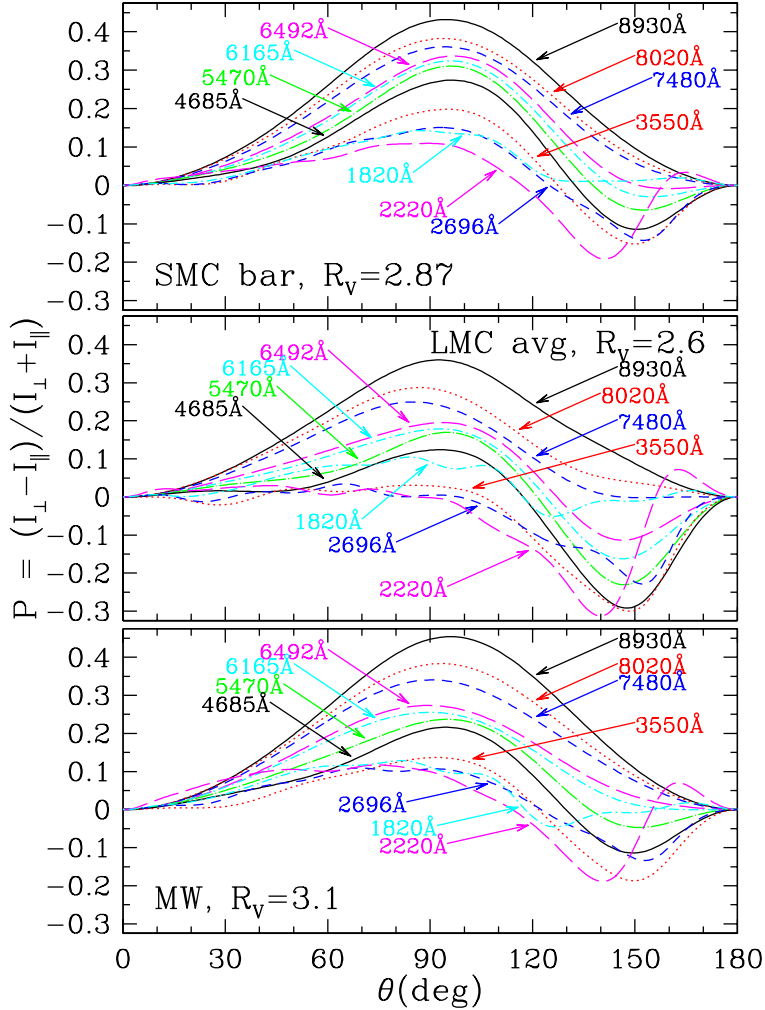}}
\caption{\footnotesize
	\label{fig:Pol}
	Degree of polarization as a function of scattering angle $\theta$,
	for scattering by Milky Way dust 
	with $R_V=3.1$, LMC average dust, and SMC bar dust.
	Curves are labelled by wavelength $\lambda$.
	}
\end{figure}

The polarization $P$ of the scattered light is shown 
in Fig.\ \ref{fig:Pol} as a function of
scattering angle $\theta$ for Milky Way dust with $R_V=3.1$, at
11 different wavelengths.
Rayleigh scattering would have 
$P=(1-\cos^2\theta)/(1+\cos^2\theta)$, with $P=1$ for $\theta=90^\circ$.
At long wavelengths, the polarization has a distinct peak near $\sim90^\circ$,
but even for $\lambda\approx 0.9\micron$ the peak polarization is only
$\sim0.45$.
As the wavelength is reduced, the peak polarization declines.
For $\lambda \ltsim 0.6\micron$, the polarization becomes negative at
large scattering angles (see Figure \ref{fig:Pol}),
with large negative polarizations in the 120--150$^\circ$ region for
$0.2\micron\ltsim \lambda  \ltsim 0.4\micron$.
Note, however, that this negative polarization occurs for scattering
angles where the scattering is very weak (see Figure \ref{fig:F_X})
and therefore could be masked by scattering at other points
on the sightline where the scattering contributes a positive
polarization.  Observations of the predicted negative polarization
will probably require simple scattering geometries, such as dust
in a thin disk, illuminated by a single source.

Similar results are found for the LMC and SMC dust mixtures -- see
Fig. \ref{fig:Pol}.
Note the very large negative polarizations found for the LMC mixture
for $0.2\micron\ltsim \lambda \ltsim 0.55\micron$.
From the variation of the ultraviolet 
polarization signature between
the different grain size distributions in Figure \ref{fig:Pol} it is
apparent that the ultraviolet polarization is sensitive to the
details of the grain size distribution.

\section{Analytic Approximations for the Phase Function
	\label{sec:phase_funcs}}
\subsection{Henyey-Greenstein Phase Function
	\label{sec:HG}}

Scattering of unpolarized incident light by a
dust mixture at wavelength $\lambda$ is characterized by the total
scattering cross section per H nucleon, $\sigma_{\rm sca}(\lambda)$, and a
``phase function''
\beq
\label{eq:phasefunc}
\Phi(\theta,\lambda) \equiv 
\frac{1}{\sigma_{\rm sca}(\lambda)} 
\frac{d\sigma_{\rm sca}(\theta,\lambda)}{d\Omega}
\eeq
characterizing the angular distribution of the scattered light,
where $d\sigma_{\rm sca}/d\Omega$ is the differential scattering
cross section, at scattering angle $\theta$, for unpolarized incident light.
The definition (\ref{eq:phasefunc}) implies the normalization
$\int d\Omega~ \Phi = 1$.
Isotropic scattering would have $\Phi=1/4\pi$.
The first moment 
\beq
\langle\cos\theta\rangle = 
\int  d\Omega \cos\theta ~\Phi(\theta,\lambda)
\eeq
of the phase function is a measure of the asymmetry between
forward and backward scattering.

Henyey \& Greenstein (1941) proposed an analytic function to
model anisotropic scattering
for dust grain mixtures:
\beq
\label{eq:HG}
\phi_0(\theta) =
\frac{1}{4\pi}
\frac{1-\gpar^2}{(1+\gpar^2-2\gpar\cos\theta)^{3/2}}
~~~,~~~ \gpar=\langle\cos\theta\rangle
~~~.
\eeq
With a single parameter $\gpar$, this is a convenient
analytic form that
has been widely used to represent dust scattering properties 
in radiative models of dusty
regions.

By construction, $\phi_{0}(\theta)$ has the correct 
first moment
$\langle\cos\theta\rangle=\int d\Omega\cos\theta~\phi_{0}(\theta)= \gpar$, 
but of course $\phi_{0}$ does not
perfectly reproduce the angular dependence of the real phase function 
$\Phi(\theta)$.
As will be seen below, $\phi_0$ is a poor approximation at both
$\lambda \ltsim 0.4\micron$ and $\lambda > 1 \micron$.

\subsection{A New Phase Function\label{sec:new_phase_func}}

At long wavelengths, Rayleigh scattering prevails, with
$\langle\cos\theta\rangle\rightarrow0$ and
$\Phi \rightarrow (3/16\pi)(1+\cos^2\theta)$.
When $\langle\cos\theta\rangle\rightarrow0$, 
the Henyey-Greenstein phase function
$\phi_{0}\rightarrow 1/4\pi$; $\phi_0$ is 33\% low at $\theta=0,\pi$ and
33\% high at $\theta=\pi/2$.

Consider the phase function
\beq
\label{eq:newPhi}
\phi_\alpha(\theta) = \frac{1}{4\pi}
\left[ \frac{1-\gpar^2}{1+\alpha(1+2\gpar^2)/3}\right] 
\frac{1+\alpha \cos^2\theta}
{(1+\gpar^2-2\gpar \cos\theta)^{3/2}}
~~~,
\eeq
with two adjustable parameters, $\alpha$ and $\gpar$. 
For $\alpha=0$, equation (\ref{eq:newPhi}) reduces to the Henyey-Greenstein
phase function $\phi_0$ (equation \ref{eq:HG}).
For $\gpar=0$ and $\alpha=1$ this reduces to the phase function for
Rayleigh scattering.
For $\alpha=1$ this corresponds to the phase function proposed by
Cornette \& Shanks (1992).
Analytic results for this phase function are given in Appendices
\ref{app:moments} - \ref{app:fixalpha}.

To determine the parameters $\alpha$ and $\gpar$ we can require
that $\phi_\alpha(\theta)$ have the same
first and second moments
$\langle\cos\theta\rangle$ and $\langle\cos^2\theta\rangle$
as $\Phi$.
The parameters $\gpar$ and $\alpha$ are then given by
equations (\ref{eq:2moment_gamma_case1}-\ref{eq:2moment_alpha}).
However,
although the resulting phase function
has correct first and second moments,
the fit is poor when the dust is strongly forward-scattering
($\langle\cos\theta\rangle\gtsim 0.6$). 

We find that an improved fit is obtained if we obtain
$\gpar$ and $\alpha$ from
equations (\ref{eq:2moment_gamma_case1}-\ref{eq:2moment_alpha})
only when the resulting $\alpha\leq1$; for values of
$\langle\cos\theta\rangle$ and $\langle\cos^2\theta\rangle$ for which
equations (\ref{eq:2moment_gamma_case1}-\ref{eq:2moment_alpha})
lead to $\alpha>1$, we instead set $\alpha=1$
and obtain $\gpar$ from equation (\ref{eq:gamma_for_fixed_alpha}).
We will refer to this new analytic phase function as $\phi_{\alpha\leq1}$.
\begin{figure}[h]
\centerline{\epsfig{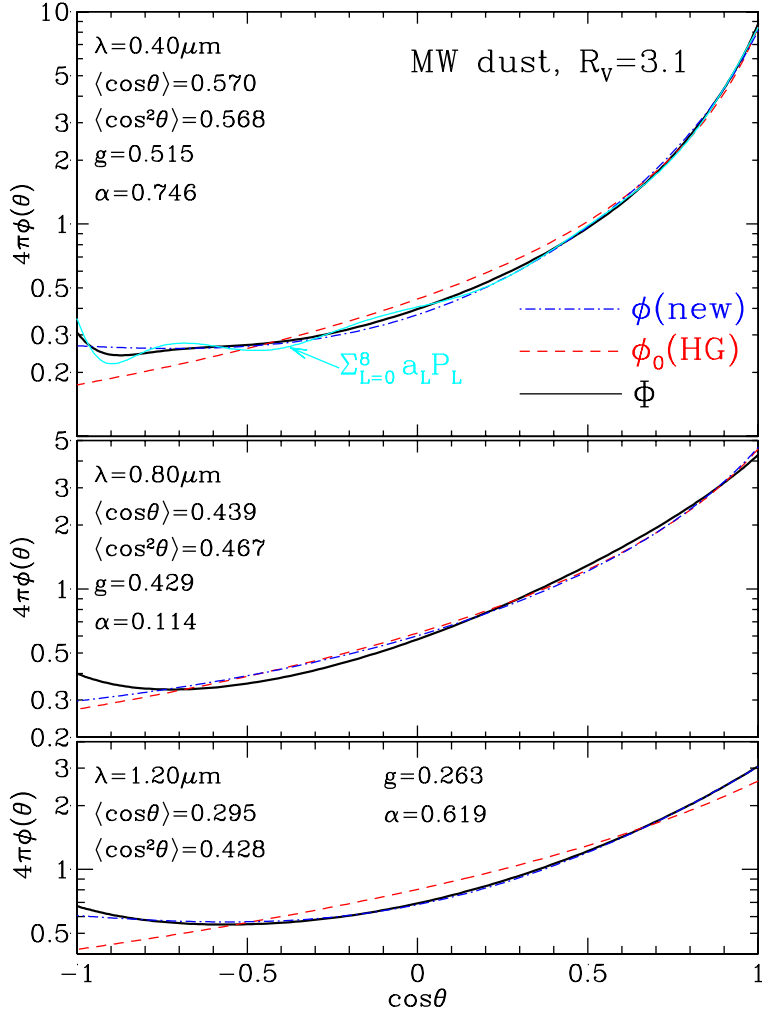}}
\caption{\footnotesize
	\label{fig:HGa}
	Scattering phase function for the WD01 Milky Way
	dust model at three wavelengths
	(solid lines) compared to the Henyey-Greenstein
	phase function $\phi_0$ and our new phase function
	$\phi_{\alpha\leq1}$.
	For $\lambda=0.4\micron$ a 9 term Legendre polynomial
	representation is also shown.
	}
\end{figure}
\begin{figure}[h]
\centerline{\epsfig{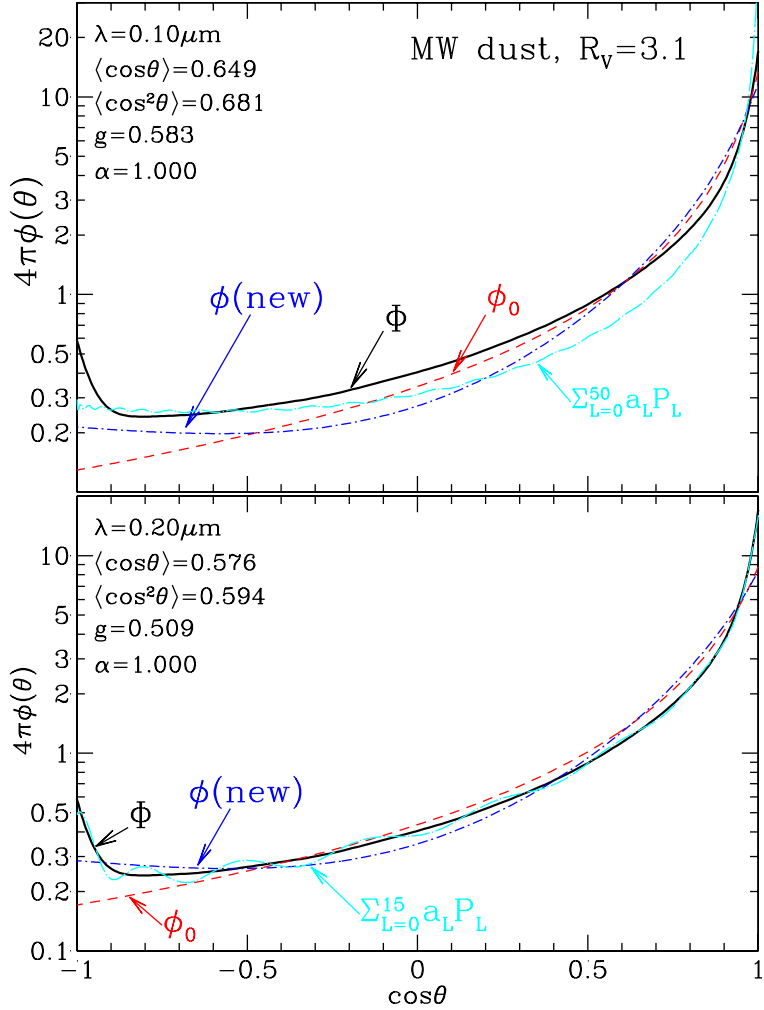}}
\caption{\footnotesize
	\label{fig:HGb}
	Same as Figure \ref{fig:HGa}, but for $\lambda=0.2\micron$
	and $0.1\micron$.
	Legendre polynomial representations of $\Phi(\theta)$
	are shown with 16 and 51 terms.
	}
\end{figure}

Figure \ref{fig:HGa} shows the scattering phase function $\Phi(\theta)$ for the
WD01 Milky Way dust model,
together with the Henyey-Greenstein phase function
$\phi_0$ and our new phase function $\phi_{\alpha\leq1}$,
at wavelengths $\lambda=1.2\micron$, $0.80\micron$, and $0.40\micron$.
For these three wavelengths, the new phase function provides
an improved fit to the actual phase function $\Phi$.
For $\lambda=0.40\micron$, we also show $\Phi$ approximated
by a sum over Legendre polynomials $P_l$ up to $l=8$.
This 9 term expansion does not reproduce $\Phi$ as
well as the
new phase function $\phi_{\alpha\leq1}$.

At ultraviolet wavelengths the grains become more forward-throwing,
and the analytic phase functions
$\phi_0$ and $\phi_{\alpha\leq1}$ no longer provide 
a good fit, as seen in Figure
\ref{fig:HGb} for $\lambda=0.20\micron$ and $0.10\micron$.
At $\lambda=0.20\micron$, for example, both the Henyey-Greenstein
phase function $\phi_0$ and the new phase function $\phi_{\alpha\leq1}$
underestimate the forward scattering intensity by a factor of $\sim$2 for
$\cos\theta \gtsim 0.98$ ($\theta\ltsim10^\circ$).

To quantify the error associated with using
an analytic phase function $\phi(\theta)$ to approximate
an actual phase function
$\Phi(\theta)$,
we define the r.m.s.\ relative error
\beq
\label{eq:hrel}
h_{\rm rel} \equiv \left[\int \frac{d\Omega}{4\pi} 
\left[\frac{\phi(\theta)-\Phi(\theta)}{\Phi(\theta)}
	\right]^2\right]^{1/2} ~~~,
\eeq
and r.m.s.\ absolute error
\beq
\label{eq:habs}
h_{\rm abs} \equiv \left[\int \frac{d\Omega}{4\pi}
\left[
\frac{\phi(\theta)-\Phi(\theta)}{\langle\Phi\rangle}
\right]^2\right]^{1/2}~~~.
\eeq
where, of course, $\langle\Phi\rangle=1/4\pi$.
The r.m.s. relative error $h_{\rm rel}$ would appear to be the best
measure of the overall quality of fit.
$h_{\rm rel}$ and $h_{\rm abs}$ will differ substantially only 
when the phase function is
very asymmetric; large fractional errors in directions where the scattering
is weak then make only a small contribution to $h_{\rm abs}$, while 
modest fractional
errors in directions where the scattering is very strong will make large
contributions to $h_{\rm abs}$.
\begin{figure}[h]
\centerline{\epsfig{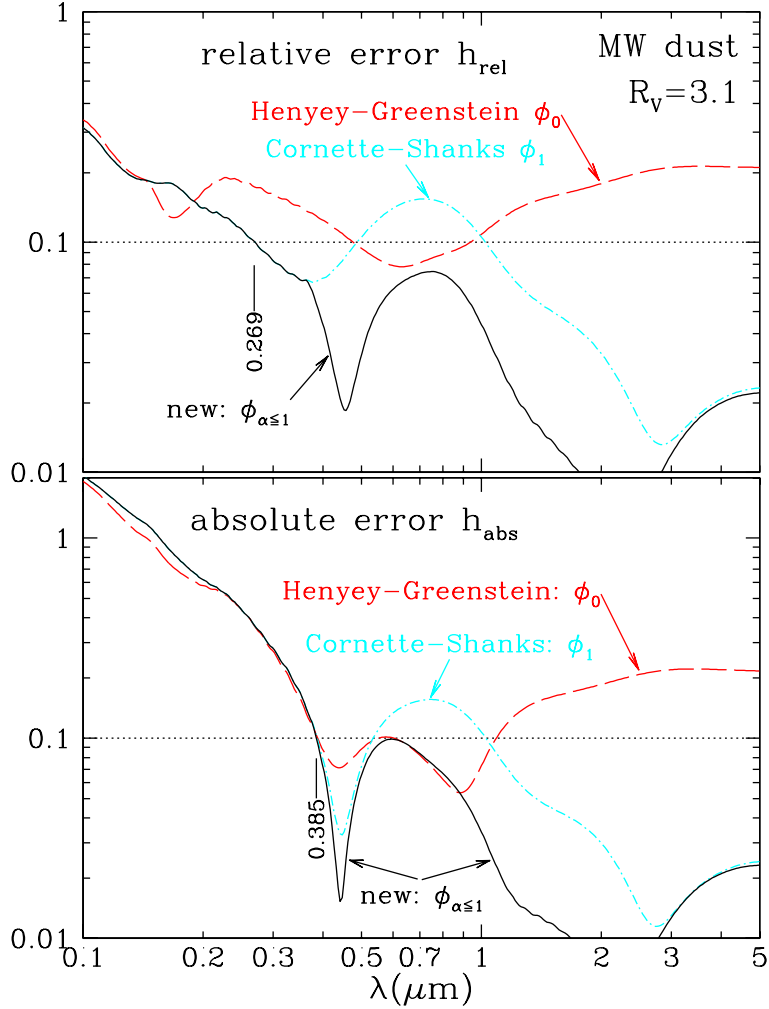}}
\caption{\footnotesize
	\label{fig:HGerr}
	Normalized relative error $h_{\rm rel}$ (eq.\ \ref{eq:hrel})
	and absolute error
	$h_{\rm abs}$ (eq.\ \ref{eq:habs})
	for the analytic phase functions $\phi_0$, $\phi_1$,
	and $\phi_{\alpha\leq1}$ applied to the 
	WD01 grain model for MW dust with $R_V=3.1$.
	}
\end{figure}

In the limit of Rayleigh scattering ($\lambda\gg a$),
$\Phi(\theta) \rightarrow \frac{3}{4}\left(1+\cos^2\theta\right)$,
and $\phi_{0}$ has errors
\begin{eqnarray}
\label{eq:h_abs}
h_{\rm abs}&\rightarrow&1/\sqrt{20}\approx0.22 ~~~,\\
\label{eq:h_rel}
h_{\rm rel}&\rightarrow&(13-4\pi)^{1/2}/3\approx 0.22 ~~~.
\end{eqnarray}

Figure \ref{fig:HGerr} compares the error of different analytic
approximations to the phase function, using as an example
$\Phi(\theta,\lambda)$ for 
the WD01 model for $R_V=3.1$ Milky Way dust.
The errors $h_{\rm rel}$ and $h_{\rm abs}$ are shown in Figure 
\ref{fig:HGerr} for the Henyey-Greenstein
function $\phi_0$,
the Cornette-Shanks phase function $\phi_1$,
and our new phase function $\phi_{\alpha\leq1}$.
At long wavelengths $\lambda\gtsim 3\micron$, the Henyey-Greenstein
approximation $\phi_0$ has
$h_{\rm rel}\approx h_{\rm abs}\approx 0.22$, as expected from
eq.\ (\ref{eq:h_abs},\ref{eq:h_rel}).
Between $\sim0.5$ and $\sim1\micron$, the relative
errors for $\phi_0$ are modest, $h_{\rm rel}<10\%$, so $\phi_0$
provides a good approximation to the actual scattering properties.
However,
$\phi_0$ has $h_{\rm rel}>10\%$ for $\lambda < 0.5\micron$, rising to
shorter wavelengths.

Figure \ref{fig:HGerr} shows the fractional errors $h_{\rm rel}$ and
$h_{\rm abs}$ for the new phase function $\phi_{\alpha\leq1}$.
As expected, the new phase function provides an excellent fit at long
wavelengths, with fractional errors $< 3\%$ for $\lambda > 1\micron$.
As the wavelength is decreased, the new phase function
$\phi_{\alpha\leq1}$ (as measured by $h_{\rm rel}$) remains preferable to
the Henyey-Greenstein phase function for $\lambda \gtsim 0.19\micron$.
$\phi_{\alpha\leq1}$ has $h_{\rm rel}<10\%$ for
$\lambda > 0.27\micron$.

In the ultraviolet the phase function becomes very strongly 
forward-scattering, and all of the
the analytic approximations have significant errors.
More complicated parameterizations -- 
such as the use of multiple Henyey-Greenstein phase functions, 
as by Witt (1977) and Hong (1985) -- may be considered.
Alternatively, a given phase function $\Phi(\theta)$ 
can be approximated by summing over
Legendre polynomials, although the sum needs to include a large number
of terms to provide a good approximation -- 16 terms are required for
$\lambda=0.20\micron$, and even 51 terms are insufficient for
$\lambda=0.10\micron$.
At short wavelengths it may be best to simply tabulate $\Phi(\theta)$
and interpolate as required.

\section{Discussion\label{sec:discuss}}

The scattering properties of interstellar dust have been determined
observationally by comparing the observed surface brightness of 
reflection nebulae with model nebulae computed with different
dust scattering properties, selecting the model which provides
the best match to the observations.
The usual approach has been to try to determine only two grain properties --
the albedo and 
$\langle\cos\theta\rangle$ -- by finding radiative transfer models which
appear to be consistent with the observed surface brightness of scattered
light.

It has been customary to assume that the scattering phase function 
$\Phi(\theta)$ can be approximated by the Henyey-Greenstein function
$\phi_0(\theta)$, with $g=\langle\cos\theta\rangle$

\begin{figure}[ht]
\centerline{\epsfig{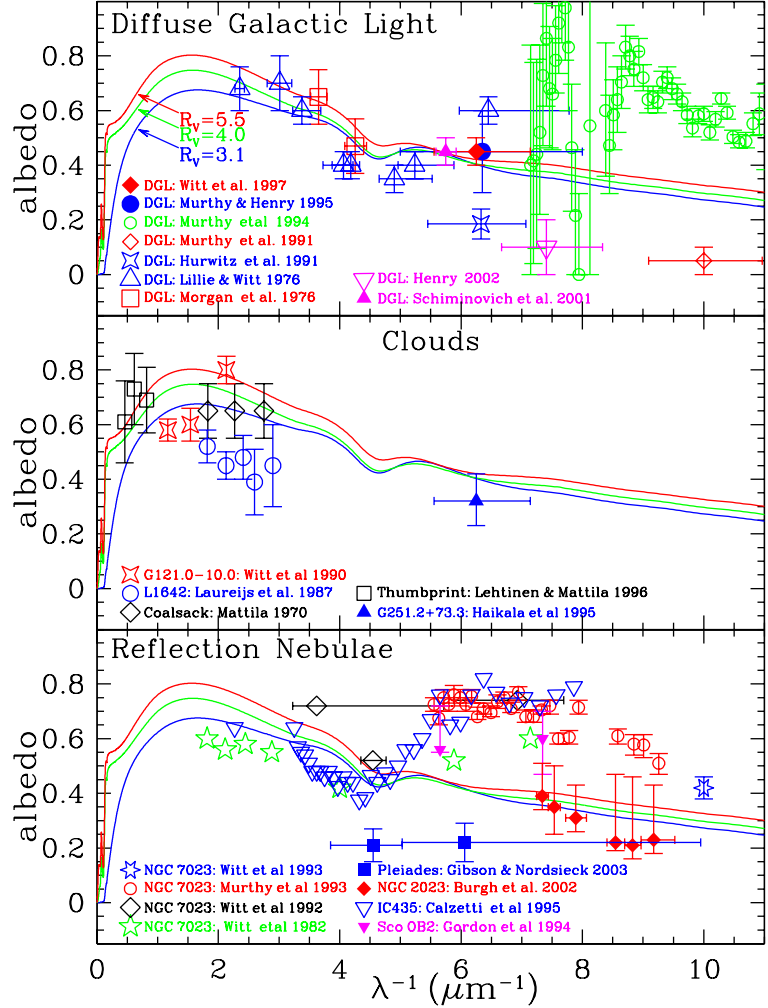}}
\caption{\footnotesize
	\label{fig:albedo}
	Scattering albedo as a function of frequency for different MW
	dust mixtures, together with observational estimates of the
	dust albedo in various regions.
	}
\end{figure}
\begin{figure}[ht]
\centerline{\epsfig{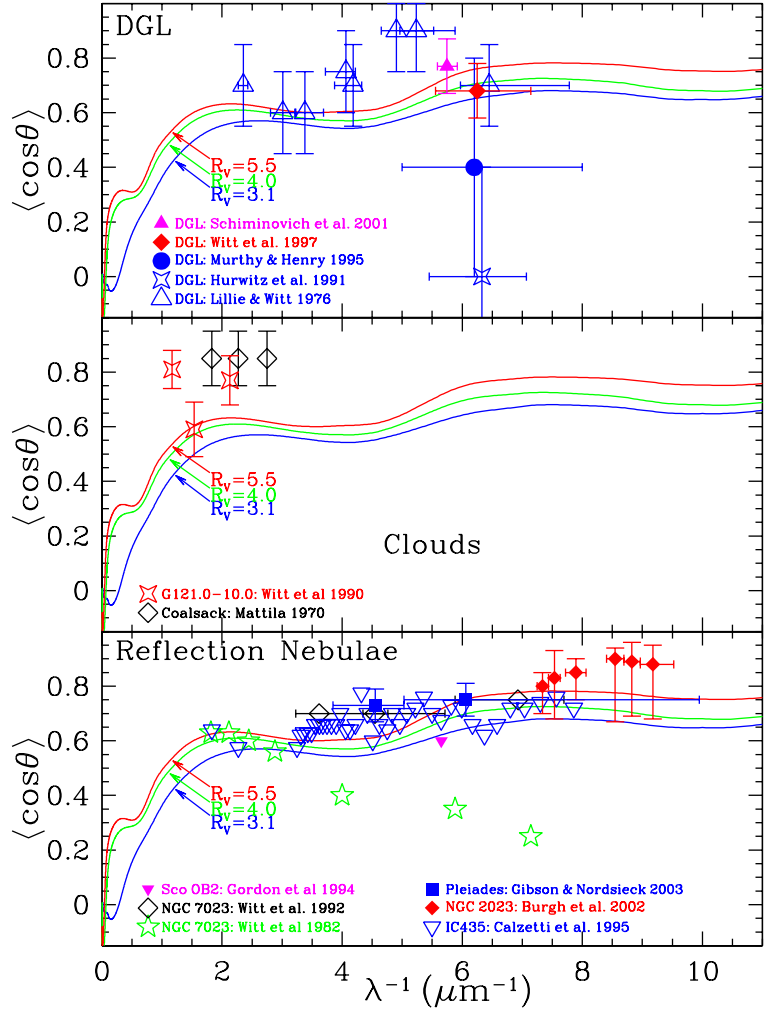}}
\caption{\footnotesize
	\label{fig:g}
	Scattering asymmetry factor $g=\langle\cos\theta\rangle$ 
	as a function of frequency for
	different MW dust mixtures.  Also shown are observational
	estimates for $g$ in various regions.
	}
\end{figure}

This approach has been used with (1) observations of high surface brightness
reflection nebulae
(e.g., NGC~2023, NGC~7023, IC~435) 
illuminated primarily by a single star,
(2) observations of individual dust clouds illuminated by ambient starlight,
and (3) observations of the much fainter diffuse galactic light
(the entire Galaxy as a reflection nebulosity).
Figures \ref{fig:albedo} and \ref{fig:g} show the results so obtained
from a number of independent studies, at wavelengths from the optical to 
vacuum ultraviolet.
Also shown are the albedo and asymmetry factor calculated for WD01 grain
models representing average Milky Way dust with $R_V\equiv A_V/E(B-V)=3.1$, 
and dust from denser regions with $R_V=4.0$ and 5.5.
While there are differences among the different observational determinations,
particularly in the ultraviolet, many of the observational results are
in general agreement with the albedo and asymmetry factor calculated for
the WD01 grain model.

For $\lambda^{-1} < 4\micron^{-1}$, the observational studies appear to
be in general agreement with one another, but at shorter wavelengths the
observational results are sometimes in conflict.
For example, from observations of the diffuse galactic light at 
$\sim6.25\micron^{-1}$ ($\lambda=0.16\micron$)
Hurwitz et al.\ (1991)
found $(a,g)=(0.185\pm.055,0.2\pm0.2)$, 
while Witt et al.\ (1997) 
found $(a,g)=(0.45\pm0.05,0.68\pm0.10)$.

Aside from genuine regional variations in dust properties, 
there are many factors which could contribute to such discrepancies.
Foremost may simply be the
difficulty of observations in the vacuum ultraviolet,
followed by uncertainties concerning the nebular geometry and the illuminating
radiation.
However, given
that use of the Henyey-Greenstein phase function $\phi_0$ introduces an
absolute error $h_{\rm abs}\gtsim70\%$ 
in the scattering phase function at $\lambda<0.16\micron$
(see Fig.\ \ref{fig:HGerr}), it seems possible that reliance on the
Henyey-Greenstein phase function may contribute to
the disagreements among different observational determinations of the
albedo and scattering asymmetry in the ultraviolet.

Unfortunately, the new phase function $\phi_{\alpha\leq1}$ is
inaccurate for $\lambda < 0.27\micron$.
For $\lambda^{-1} \gtsim 4\micron^{-1}$, radiative transfer models
should use a more accurate representation of the phase function $\Phi$.
As discussed above, one possibility is expansion in Legendre
polynomials, although many terms are required.
Alternatively, it may be simplest to use
tabulated values of $\Phi$ with interpolation.
``Inversion'' of the observational data to infer the grain properties
may not be feasible; the best approach may be to assume the dust properties
[i.e., $\sigma_{\rm ext}(\lambda)$, albedo($\lambda$), 
and $\Phi(\theta,\lambda)$] based on a grain model,
and then seek a dust spatial distribution which maximizes consistency
with the observed surface brightness and whatever other constraints
are available.  Failure to find agreement would suggest that the
grain model may be incorrect.

Applying this approach to reflection nebulae is
complicated by the possibility of clumpy dust distributions, which
Mathis et al.\ (2002) show can produce a 
range of ratios of nebular surface brightness to unscattered stellar flux,
depending on the detailed dust clump distribution and the viewing direction.
Diffuse clouds (i.e., the diffuse galactic light) and externally-illuminated
clouds may prove easier to interpret than bright reflection nebulae,
which generally have a bright star embedded in the dust distribution.

\section{Summary\label{sec:summary}}

The following are the principal results of this work:
\begin{enumerate}

\item Differential scattering cross sections
have been calculated at selected wavelengths 
from infrared to ultraviolet for dust mixtures appropriate to the
diffuse interstellar medium in the Milky Way, the LMC, and the SMC.
These scattering functions, which can be used for modeling reflection nebulae,
are available at http://www.astro.princeton.edu/$\sim$draine .

\item The polarization of scattered light, as a function of scattering angle,
is calculated at selected wavelengths.

\item The calculated phase function $\Phi(\theta)$
for the WD01 model for $R_V=3.1$ MW dust is compared to the widely-used
Henyey-Greenstein phase function $\phi_0(\theta)$.
At wavelengths $0.48\micron<\lambda<0.96\micron$,
$\phi_0$ provides a good fit to $\Phi$,
with r.m.s. relative error $h_{\rm rel}<10\%$.
In the ultraviolet, however, $\phi_0$ provides
a poor fit, with r.m.s. absolute error
$h_{\rm abs} > 50\%$ at $\lambda < 2400\Angstrom$.
The Henyey-Greenstein phase function is not suitable for
accurate modeling of reflection nebulae in the ultraviolet.

\item A new phase function, $\phi_{\alpha\leq1}(\theta)$ is proposed
[see equation (\ref{eq:newPhi})].
This phase function provides a good approximation to our calculated phase
functions for $\lambda > 0.27\micron$, with r.m.s. relative error
$h_{\rm rel}< 10\%$.
At shorter wavelengths, however, $\Phi$ becomes
strongly forward-throwing, and neither the Henyey-Greeenstein phase function
$\phi_0$ nor the new phase function $\phi_{\alpha\leq 1}$
provide good approximations.

\item There are significant discrepancies among 
observational determinations of the
albedo and $\langle\cos\theta\rangle$ in the ultraviolet
for the diffuse galactic light
and dust in discrete clouds and reflection nebulae.
One possible cause for these discrepancies may be 
reliance on the Henyey-Greenstein
phase function in the radiative transfer models.
It is recommended that the new phase function $\phi_{\alpha\leq1}$ be used
at wavelengths $\lambda \gtsim 0.27\micron$; at shorter wavelengths
accurate radiative transfer models should use some other procedure to
obtain $\Phi$.
\end{enumerate}
\acknowledgements

I thank
Karl Gordon, Aigen Li,
Kalevi Mattila,
Jonathan Tan, Adolf Witt, and Michael Woolf for valuable comments,
the anonymous referee for helpful suggestions,
and Robert Lupton for making available the SM software package.
This work was supported in part by
NSF grant AST-9988126.

\appendix

\section{Phase Function Moments\label{app:moments}}

A phase function of the form
\beq
\phi_\alpha(\theta) = \frac{1}{4\pi}
\left[\frac{1-\gpar^2}{1+\alpha(1+2\gpar^2)/3}\right]
\frac{1+\alpha\cos^2\theta}{(1+\gpar^2-2\gpar\cos\theta)^{3/2}}
\eeq
has first and second moments given by
\begin{eqnarray}
\label{eq:1st_moment}
\langle\cos\theta\rangle &=&
\gpar 
\frac{1+\alpha(3+2\gpar^2)/5}{1+\alpha(1+2\gpar^2)/3}
~~~,
\\
\label{eq:2nd_moment}
\langle\cos^2\theta\rangle &=&
\frac{1+2\gpar^2+(3\alpha/35)
\left(7+20\gpar^2+8\gpar^4\right)}
{3+\alpha(1+2\gpar^2)}
~~~.
\end{eqnarray}

\section{Reproduce $\langle\cos\theta\rangle$ and 
$\langle\cos^2\theta\rangle$\label{app:fit2moms}}

If we specify $\langle\cos\theta\rangle$ and $\langle\cos^2\theta\rangle$,
the adjustable parameters $\gpar$ and $\alpha$ are determined
from equations (\ref{eq:1st_moment},\ref{eq:2nd_moment}).
It can be shown that $\gpar$ satisfies the cubic equation
\begin{equation}
\gpar^3 -\frac{17}{3}\langle\cos\theta\rangle \gpar^2
+ 7\langle\cos^2\theta\rangle \gpar - \frac{7}{3}\langle\cos\theta\rangle = 0
~~~.
\end{equation}
Let
\begin{eqnarray}
a&\equiv& \frac{7}{3}\langle\cos^2\theta\rangle -\frac{289}{81}\langle\cos\theta\rangle^2
~~~,
\\
b&\equiv& 
\frac{119}{18}\langle\cos\theta\rangle\langle\cos^2\theta\rangle
-\frac{4913}{729}\langle\cos\theta\rangle^3
-\frac{7}{6}\langle\cos\theta\rangle
~~~.
\end{eqnarray}
If $a^3+b^2>0$, the solution is
\beq
\label{eq:2moment_gamma_case1}
\gpar =
\left[\left(a^3+b^2\right)^{1/2}-b\right]^{1/3}
-
\left[\left(a^3+b^2\right)^{1/2}+b\right]^{1/3}
+\frac{17}{9}\langle\cos\theta\rangle
~~~,
\eeq
whereas if $a^3+b^2<0$, the appropriate root is
\beq
\label{eq:2moment_gamma_case2}
\gpar = 2|a|^{1/2}\cos(\psi/3) + \frac{17}{9}\cos\theta
~~~~~,~~~~~
\psi \equiv \arccos\left(-b/|a|^{3/2}\right)
~~~.
\eeq
The parameter $\alpha$ is then obtained from equation (\ref{eq:1st_moment}):
\beq
\label{eq:2moment_alpha}
\alpha = \frac{25(\langle\cos\theta\rangle-\gpar)}
{3(3+2\gpar^2)\gpar-5(1+2\gpar^2)\langle\cos\theta\rangle}
~~~.
\eeq

\section{Fix $\alpha$, Reproduce $\langle\cos\theta\rangle$
	\label{app:fixalpha}}

If we choose to fix the value of $\alpha$ and
$\langle\cos\theta\rangle$, then 
$\gpar$ must satisfy
\beq
\label{eq:cubic_for_fixed_alpha}
\gpar^3 -\frac{5}{3}\langle\cos\theta\rangle\gpar^2
+ \left(\frac{3\alpha+5}{2\alpha}\right)\gpar - 
\frac{5}{6\alpha}(3+\alpha)\langle\cos\theta\rangle
= 0 ~~~.
\eeq
For $\alpha\neq0$, equation (\ref{eq:cubic_for_fixed_alpha})
has the real root
\begin{eqnarray}
\label{eq:gamma_for_fixed_alpha}
\gpar &=& \left[(a_\alpha^3+b_\alpha^2)^{1/2}+b_\alpha\right]^{1/3}
- \left[(a_\alpha^3+b_\alpha^2)^{1/2}-b_\alpha\right]^{1/3} + 
\frac{5}{9}\langle\cos\theta\rangle
~~~,
\\
a_\alpha &\equiv& \frac{1}{2}+
	\frac{5}{6\alpha}-
	\frac{25}{81}\langle\cos\theta\rangle^2
~~~,
\\
b_\alpha &\equiv& \frac{125}{729}\langle\cos\theta\rangle^3+
	\frac{5}{9\alpha}\langle\cos\theta\rangle
~~~.
\end{eqnarray}
For $\alpha\rightarrow0$ (the Henyey-Greenstein phase function)
we see immediately that equation (\ref{eq:cubic_for_fixed_alpha})
requires $\gpar=\langle\cos\theta\rangle$.



\begin{thebibliography}{}



\bibitem[Bohren \& Huffman 1984]{BH84}
	Bohren, C.F., \& Huffman, D.R. 1983,
	Absorption and Scattering of Light by Small Particles
	(New York: Wiley)




\bibitem[Burgh et al 2002]{BMF02}
	Burgh, E.B., McCandliss, S.R., \& Feldman, P.D.
	2002,
	ApJ, 575, 240


\bibitem[Calzetti etal 1995]{CBG95}
	Calzetti, D., Bohlin, R.C., Gordon, K.D., Witt, A.N., \& 
	Bianchi, L.
	1995,
	ApJ, 446, L97

\bibitem[Cornette \& Shanks 1992]{CS92}
	Cornette, W.M., \& Shanks, J.G. 1992,
	Appl. Optics, 31, 3152


\bibitem[Draine 2003a]{Dr03a}
	%
	Draine, B.T. 2003a,
	ARAA, 41, 241

\bibitem[Draine 2003b]{Dr03b}
	Draine, B.T. 2003b,
	ApJ, accepted 
	(Paper II)
	astro-ph/0308251

\bibitem[Draine \& Lee 1984]{DL84} 
	Draine, B.T., \& Lee, H.-M. 1984, ApJ, 468, 269 (DL84)


\bibitem[Draine \& Malhotra 1993]{DM93}
	Draine, B.T., \& Malhotra, S. 1993,
	ApJ, 414, 632

\bibitem[Draine \& Tan 2003]{DT03}
	Draine, B.T., \& Tan, J.C. 2003,
	ApJ, 594, 000 (Sept 1)
	[http://arxiv.org/abs/astroph/0208302]

\bibitem[Dufour 1984]{Du84}
	Dufour, R.J. 1984,
	in IAU Symp. 108, Structure and Evolution of the Magellanic Clouds,
	ed. S. van den Bergh \& K.S. de Boer (Dordrecht: Reidel), 353


\bibitem[Fitzpatrick 1986]{Fi86}
	Fitzpatrick, E.L. 1986,
	AJ, 92, 1068



\bibitem[Gibson \& Nordsieck 2003]{GN03}
	Gibson, S.J., \& Nordsieck, K.H. 2003,
	ApJ, 589, 362


\bibitem[Gordon et al 1994]{GWC94}
	Gordon, K.D., Witt, A.N., Carruthers, G.R., Christensen, S.A.,
	Dohne, B.C. 1994,
	ApJ, 432, 641

\bibitem[Haikala et al. 1995]{HMB95}
	Haikala, L.K., Mattila, K., Bowyer, S., Sasseen, T.P.,
	Lampton, M., \& Knude, J. 1995,
	ApJ, 443, L33




\bibitem[Henry 2002]{He02}
	Henry, R.C. 2002,
	ApJ, 570, 697

\bibitem[Henyey \& Greenstein 1941]{HG41}
	Henyey, L.G., \& Greenstein, J.L. 1941,
	ApJ, 93, 70

\bibitem[Hong 1985]{Ho85}
	Hong, S.S. 1985,
	A\&A, 146, 67

\bibitem[Hurwitz et al 1991]{HBM91}
	Hurwitz, M., Bowyer, S., \& Martin, C.
	1991,
	ApJ, 372, 167

\bibitem[Koorneef 1982]{Ko82}
	Koorneef, J. 1982,
	A\&A, 107, 247


\bibitem[Kurt \& Dufour 1998]{KD98}
	Kurt, C.M., \& Dufour, R.J. 1998,
	Rev. Mex. Astr. Ap. 7, 202

	

\bibitem[Laureijs et al 1987]{LMS87}
	Laureijs, R.J., Mattila, K., \& Schnur, G. 1987,
	A\&A, 184, 269

\bibitem[Lehtinen \& Mattila 1996]{LM96}
	Lehtinen, K., \& Mattila, K. 1996,
	A\&A, 309, 570


\bibitem[Li \& Draine 2001]{LD01}
	Li, A., \& Draine, B.T. 2001,
	ApJ, 554, 778

\bibitem[Li \& Draine 2002]{LD02}
	Li, A., \& Draine, B.T. 2002,
	ApJ, 572, 232

\bibitem[Lillie \& Witt 1976]{LW76}
	Lillie, C.F., \& Witt, A.N.
	1976,
	ApJ, 208, 64



\bibitem[Martin et al 1989]{MML89}
	Martin, N, Maurice, E., \& Lequeux, J. 1989,
	AA, 215, 129



\bibitem[Mathis et al 2002]{MWW02}
	Mathis, J.S., Whitney, B.A., \& Wood, K. 2002,
	ApJ, 574, 812

\bibitem[Mattila 1970]{Mat70}
	Mattila, K. 1970,
	A\&A, 9, 53



\bibitem[Morgan etal 1976]{MKT76}
	Morgan, D.H., Nandy, K., \& Thompson, G.I.
	1976,
	MNRAS, 177, 531

\bibitem[Murthy et al 1993]{MDH93}
	Murthy, J., Dring, A., Henry, R.C., Kruk, J.W., Blair, W.P.,
	Kimble, R.A., \& Durrance, S.T. 1993,
	ApJ, 408, L97

\bibitem[Murthy \& Henry 1995]{MH95}
	Murthy, J., \& Henry, R.C.
	1995,
	ApJ, 448, 848

\bibitem[Murthy et al 1991]{MHH91}
	Murthy, J., Henry, R.C., \& Holberg, J.B. 1991,
	ApJ, 383, 198

\bibitem[Murthy et al 1994]{MHH94}
	Murthy, J., Henry, R.C., \& Holberg, J.B. 1994,
	ApJ, 428, 233




\bibitem[Pendleton \& Allamandola 2002]{PA02}
	Pendleton, Y.J., \& Allamandola, L.J. 2002,
	ApJS, 138, 75




\bibitem[Schiminovich etal 2001]{SFM01}
	Schiminovich, D., Friedman, P.G., Martin, C., \& Morrissey, P.F.
	2001,
	ApJ, 563, L161
















\bibitem[Weingartner \& Draine 2001]{WD01}
	Weingartner, J.C., \& Draine, B.T. 2001,
	ApJ, 548, 296 (WD01)

\bibitem[Wiscombe 1980]{Wi80}
	Wiscombe, W.J. 1980,
	Appl. Opt., 19, 1505

\bibitem[Wiscombe 1996]{Wi96}
	Wiscombe, W.J. 1996,
	NCAR Technical Note NCAR/TN-140+STR,\\
	ftp://climate.gsfc/nasa.gov/pub/wiscombe/SingleScatt/Homogen\_Sphere/Exact\_Mie/
	NCARMieReport.pdf

\bibitem[Witt 1977]{Wi77}
	Witt, A.N. 1977,
	ApJS, 35, 1

\bibitem[Witt etal 1997]{WFS97}
	Witt, A.N., Friedmann, B.C., \& Sasseen, T.P.
	1997,
	ApJ, 481, 809

\bibitem[Witt etal 1990]{WOS90}
	Witt, A.N., Oliveri, M.V., \& Schild, R.E. 1990,
	AJ, 99, 888

\bibitem[Witt etal 1992]{WBP92}
	Witt, A.N., Petersohn, J.K., Bohlin, R.C., O'Connell, R.W., 
	Roberts, M.S.,
	et al.,
	1992,
	ApJ, 395, L5

\bibitem[Witt et al 1993]{WPH93}
	Witt, A.N., Petersohn, J.K., Holberg, J.B., Murthy, J., Dring, A.,
	\& Henry, R.C. 1993,
	ApJ, 410, 714

\bibitem[Witt et al 1982]{WWB82}
	Witt, A.N., Walker, G.A.H., Bohlin, R.C., \& Stecher, T.P. 1982,
	ApJ, 261, 492





\end{thebibliography}
\end{document}